\documentclass[sigconf]{acmart}

\usepackage[ruled, vlined, linesnumbered]{algorithm2e}
\usepackage[caption=false]{subfig}
\usepackage{tikz}

\usepackage{hyperref} 

\usepackage{multirow}

\newtheorem{definition}{Definition}
\newtheorem{theorem}{Theorem}
\newtheorem{lemma}{Lemma}

\author{Martin Byrenheid}
\affiliation{TU Dresden}
\email{martin.byrenheid@tu-dresden.de}

\author{Stefanie Roos}
\affiliation{Delft University of Technology}
\email{s.roos@tudelft.nl}

\author{Thorsten Strufe}
\affiliation{KIT Karlsruhe}
\email{thorsten.strufe@kit.edu}

\graphicspath{{./images/}}
 
\begin{document}

\title{Topology Inference of Networks utilizing Rooted Spanning Tree Embeddings}

\begin{abstract}
Due to its high efficiency, routing based on greedy embeddings of rooted spanning trees is a promising approach for dynamic, large-scale networks with restricted topologies.
Friend-to-friend (F2F) overlays, one key application of embedding-based routing, aim to prevent disclosure of their participants to malicious members by restricting exchange of messages to mutually trusted nodes. 
Since embeddings assign a unique integer vector to each node that encodes its position in a spanning tree of the overlay, attackers can infer network structure from knowledge about assigned vectors.
As this information can be used to identify participants, an evaluation of the scale of leakage is needed.

In this work, we analyze in detail which information malicious participants can infer from knowledge about assigned vectors.
Also, we show that by monitoring packet trajectories, malicious participants cannot unambiguously infer links between nodes of unidentified participants. 
Using simulation, we find that the vector assignment procedure has a strong impact on the feasibility of inference. 
In F2F overlay networks, using vectors of randomly chosen numbers for routing decreases the mean number of discovered individuals by one order of magnitude compared to the popular approach of using child enumeration indexes as vector elements.
\end{abstract}

\maketitle

\section{Introduction}
\label{sec:intro}
Embedding-based routing algorithms rely on the assignment of a distinct logical coordinate to every node in a network.
To discover routes between nodes, such routing algorithms rely on a metric function that indicates the logical distance between coordinates.

A highly promising approach for large-scale networks with low diameter are rooted spanning tree embeddings~\cite{herzen2011scalable,chavez2007routing,houthooft2015robust,roos2016anonymous}, where the logical coordinate of each node is an integer vector that uniquely encodes its position in a rooted spanning tree over the network.
Such embeddings enable routing with low path stretch while requiring each node to keep only a polylogarithmic number of bits per neighbor as routing information~\cite{herzen2011scalable}.
Furthermore, multiple rooted trees can be leveraged in parallel to enable routing despite intermittent failures~\cite{houthooft2015robust,roos2016anonymous}.

Due to its high efficiency, routing based on rooted spanning tree embeddings is well-suited for Friend-to-Friend (F2F) overlay networks, such as the dark Freenet~\cite{clarke2010private} or GNUnet~\cite{grothoff2017}'s friend-to-friend mode.
These overlays restrict connectivity to mutually trusted nodes to achieve strong security and privacy in the presence of malicious participants.
To set up connections to nodes of other participants, an attacker needs to perform social engineering, which we consider to be costly to conduct on a large scale.

One of the key properties F2F overlays aim to achieve is \emph{membership concealment}~\cite{vasserman2009membership}: identifying information, such as the IP address of a node, is not revealed to any untrusted participants.
Here, these networks differ dramatically from anonymity networks such as Tor, which reveal the IP address to the guard or bridge node~\cite{dingledine2004tor}. 
However, due to the trust-based restriction of connectivity, the structure of a F2F overlay resembles the social graph of its participants.
Previous studies have shown that unknown individuals in a social graph can be de-anonymized by looking for nodes with similar structural properties in another, non-anonymous social graph~\cite{narayanan2009anonymizing,korula2014efficient} obtained from publicly available data, e.g., by crawling online social networks.
As a consequence, distributed algorithms that operate on F2F overlays, such as routing, should minimize exposure of overlay structure.

As the logical coordinate of each node $u$ in a rooted spanning tree embedding corresponds to a path from $u$ to the root of the spanning tree, this routing approach inherently leaks information about the structure of the encoded spanning tree and thus also the overlay structure.
Furthermore, as this approach also leverages non-tree links during routing, colluding participants may be able to obtain additional information about links between overlay nodes by tracking which nodes a message has traversed, which makes de-anonymization attacks more accurate.
Consequently, the usage of rooted spanning tree embeddings conflicts with the aforementioned goal of membership concealment.
Yet, there is no work that quantifies the actual privacy loss caused by the logical coordinates.

While topology-hiding communication protocols have been proposed in the literature, they either rely on flooding for route discovery~\cite{zhang2012design} or perform broadcast to all participants for each message~\cite{akavia2020topology} and thus incur prohibitively high overhead for communication in large networks.
Thus, such protocols do not pose a suitable alternative to embedding-based routing. 

In this paper, we present the following contributions:
\begin{itemize}
   \item We formalize the concept of topological knowledge about an overlay network and explain in detail which knowledge an attacker can infer from observed logical coordinates of a rooted spanning tree embedding.
   \item We show that if data messages do not carry the logical coordinate of their originator in plain text, then colluding malicious participants cannot unambiguously infer links incident to nodes beyond their direct neighborhood.
 \item We perform an extensive simulation study for two state of the art algorithms to evaluate the number of previously unknown participants that malicious participants can infer from logical coordinates propagated by the embedding algorithms. 
\end{itemize}

\noindent The results of our simulation study show that in social graph-like overlay networks, the way logical coordinates are assigned has a strong impact on the number of participants that can be discovered. If coordinate elements are determined by enumeration of child nodes, an attacker can infer roughly one order of magnitude more participants than if vectors of random numbers are used as coordinates.

\section{Related Work}
\label{sec:related-work}
While there are no studies on the inference of topology from embed\-ding-based routing, the inference of network structure from other routing algorithms, in particular IP routing, has been addressed several times. In the following, we thus give an overview of state-of-the-art methods in the context of IP routing and discuss their applicability to embedding-based routing for F2F overlays.

One of the first approaches to obtain a snapshot of the Internet was by means of sending IP packets with varying initial values in their Time-To-Live (TTL) field~\cite{spring2004measuring,donnet2005improved,holbert2015network,jin2008scalable}.
Whenever the TTL of an IP packet reaches zero during transit, many Internet routers drop the packet and send a notification towards the originator of the message.
As the notification contains the IP address of the reporting router, paths between different endpoints can be recovered by sending packets with increasing initial TTL values between them while recording the received notification messages.
Embedding-based routing schemes for F2F overlays do not have such a notification mechanism, so similar approaches for exploring the topology are not applicable.

Works from the area of \emph{network tomography} infer the topology between multiple nodes based on end-to-end probe measurements of network characteristics, such as message loss or delay~\cite{ni2009efficient,krishnamurthy2012robust,malekzadeh2013network,coates2002maximum}.
If there is a high correlation between two nodes $u$ and $v$ when probes are sent by the same node $n$, then it is assumed that the path from $n$ to $u$ overlaps with the path from $n$ to $v$ and thus, there must be a common node $w$ on both paths.

However, tomography can detect if paths are likely to overlap but cannot reveal the number of overlapping nodes or the actual length of the paths.
Thus, the inferred topology may contain fewer nodes than there actually are.
To overcome this limitation, network tomography approaches have been combined with notification messages~\cite{ni2009efficient} or packets with a limited hop number~\cite{malekzadeh2013network}.
As explained before, embedding-based routing schemes for F2F overlays do not provide packet loss notification mechanisms. 
As routing on greedy embeddings does not suffer from routing loops, limiting the maximum number of hops is furthermore unnecessary. 

In settings where nodes can learn their hop distance to all other nodes in the network, estimates on network topologies can be inferred from the hop distances of a subset of nodes~\cite{bouchoucha2019topology}.
So far, no existing F2F network supports collection of hop distance from one node to all other nodes.
However, when embeddings based on breadth-first-search spanning trees are used for routing, every node can learn its hop distance to a subset $S$ of nodes from the logical coordinate of their neighbors.
The algorithm of Bouchoucha et al.~\cite{bouchoucha2019topology} then enables inference of links between nodes in $S$.

However, performing topology inference solely from hop distance information disregards further information that is available to the adversary.
For example, if the adversary discovers two nodes $u$ and $v$ that are two hops away from one of his nodes and one node $w$ that is three hops away, he cannot tell if $w$ is connected to $u$ or to $v$.
We will show in Section~\ref{sec:leakage} that adversaries can easily infer some of those links between nodes in $S$ from the logical coordinates.
Furthermore, all links that are inferred by prior algorithm are also included by the inference attacks presented in Section~\ref{sec:leakage}.
Thus, our algorithm is able to infer more topological information in networks that use embedding-based routing.

\section{System model}
\label{sec:model}
\noindent In the following, we explain our system model, including our terminology, and subsequently state our adversary model.

\subsection{Network Model}
\label{sec:network-model}
We consider overlay networks with bidirectional connections and thus model an overlay as an undirected graph $O = (V,E)$, where $V$ presents the set of participating nodes and an edge represents a connection between two nodes.
We say that $u \in V$ is a neighbor of $v \in V$ iff $(u,v) \in E$.
In the following, we define the neighborhood of a set of nodes $V' \subset V$ in $O$ as $N_O(V') = \{u \mid u \in V \setminus V',\exists v \in V': (u,v) \in E \}$.

We do not assume that participating nodes have knowledge about the structure of the overlay beyond their direct neighborhood.

\paragraph{Embedding-based routing}
To enable communication between nodes that are not neighbors in the overlay, the network leverages routing based on rooted spanning tree embeddings~\cite{herzen2011scalable,chavez2007routing}.
In these embeddings, a unique vector of integers $c \in \mathbb{N}_0^*$ is assigned to every node that encodes its position in a rooted spanning tree over the network.
Each such vector then denotes the \emph{logical coordinate} of the corresponding node in a virtual space.
To do so, state-of-the-art distributed embedding algorithms~\cite{hoefer13greedy,herzen2011scalable,houthooft2015robust} first form a rooted spanning tree over the current overlay.
Afterwards, the root node $r$ of the spanning tree sets a predefined vector $c_r$ as its logical coordinate.
Subsequently, node $r$ determines an ordering $v_0,v_1,\ldots,v_{|N_O(r)|}$ among its children in the spanning tree and assigns the vector $c_r || i$ to the $i$-th child for each $i \in \{0,\ldots,|N_O(r)|\}$, where "||" denotes concatenation.
As soon as a child $u$ of $r$ has set its logical coordinate $c_u$ accordingly, it analogously determines an order among its children and assigns $c_u || i$ to its $i$-th child.
This process continues until every node has obtained a logical coordinate.

\begin{figure}
   \centering
  \subfloat[][Initial overlay state]{
   \def\svgwidth{0.275\columnwidth}
   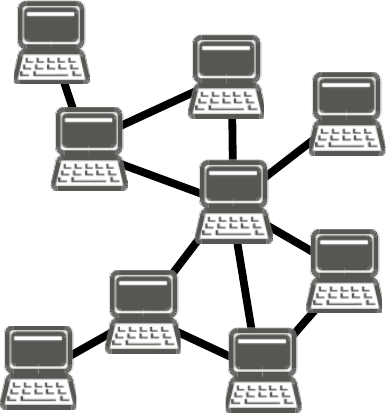
}
\hspace{0.5em}
  \subfloat[][Rooted tree formation]{
   \def\svgwidth{0.275\columnwidth}
   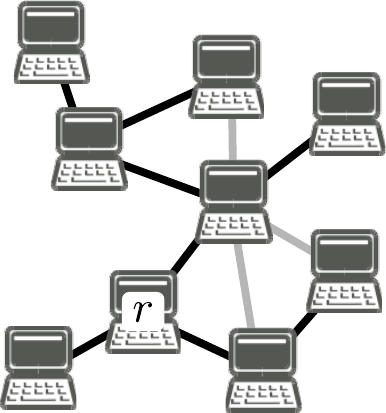
}
\hspace{0.5em}
  \subfloat[][Coordinate assignment]{
   \def\svgwidth{0.275\columnwidth}
   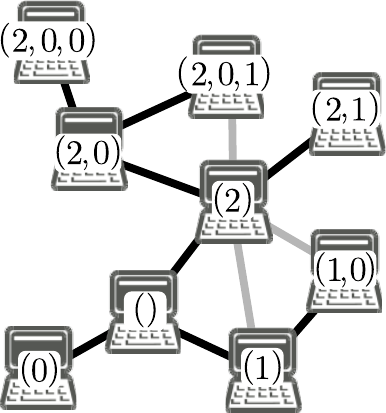
}
   \caption{Example for coordinate assignment produced by a rooted spanning tree embedding. Grey edges denote non-tree links.}
   \label{fig:example-embedding}
\end{figure}

For simplicity, in this paper we assume that the empty vector is assigned to the root node, i.e. $c_r = ()$, as proposed by Höfer et al.~\cite{hoefer13greedy}.
The attacks presented in the following sections can however easily be adapted for other root coordinate assignments.
Figure~\ref{fig:example-embedding} shows an example for such an assignment of logical coordinates.

In the following, we say that a node $u$ with coordinate $c_u$ is the \emph{parent} of a node $v$ with coordinate $c_v$ if $c_u$ is a prefix of $c_v$ and $|c_u| = |c_v|-1$.
In this case, we also say $v$ is a \emph{child} of $u$.
We say that $u$ is a \emph{sibling} of $v$ if both have the same parent.
Furthermore, we say that $v$ is a \emph{descendant} of $u$ if $c_u$ is a prefix of $c_v$.

For the actual routing of messages, nodes determine the logical distance between coordinates by means of the \emph{tree distance} 
\begin{equation}
   \delta(c_a,c_b) = |c_a| \cdot |c_b| - 2cpl(c_a,c_b) 
\end{equation}
where "$|c|$" denotes the length of the vector $c$ and "$cpl(c_a,c_b)$" denotes the length of the longest common prefix of $c_a$ and $c_b$.
When a node $u$ receives a message with target coordinate $c_t$ that differs from the coordinate assigned to $u$, $u$ forwards the message to a neighbor $v$ with coordinate $c_v$ such that $\delta_{TD}(c_v,c_t) < \delta_{TD}(c_u,c_t)$.

An important feature of the embeddings considered here is that during routing, nodes check the coordinates of \emph{all} their neighbors, including those that are neither their parent nor their child.
Latter property allows routing to find shorter paths than those found by simple spanning tree routing~\cite{herzen2011scalable} and allows the discovery of alternate paths in case of failures~\cite{roos2016anonymous,houthooft2015robust}.
In the following, we denote overlay links that are part of the spanning tree as \emph{tree links} while all other links are called \emph{shortcut links}.

\subsection{Adversary model}
F2F overlays such as Freenet~\cite{clarke2010private} offer services like messaging and publishing in a censorship-resistant and anonymous manner, making it a valuable communication tool for journalists and activists.
In this work, we therefore consider a malicious actor that aims to identify the participants of a F2F network, e.g., to uncover activist networks.

Due to the trust-based formation of links, the topology of F2F overlays corresponds to the graph of mutual acquaintances between its participants.
It therefore seems likely that the F2F overlay topology resembles other graphs that represent social interactions and relationships, such as those obtained from crawling online social networks or phone call records~\cite{narayanan2009anonymizing}.
If the attacker is able to infer a subgraph $O'=(V',E')$ of the overlay, they\footnote{We refer to the attacker using the singular they~\cite{apa_style}.} can then leverage graph-based de-anonymization attacks~\cite{narayanan2009anonymizing,yartseva2013performance,korula2014efficient,sharad2016change} to infer the identity of node operators.
Such de-anonymization attacks heuristically find mappings between the nodes of two graphs based on structural features, such as common neighbors or node degrees.
The adversary thus aims to infer as much information as possible about the overlay graph $O$ to increase the number of mappings that can be found and to increase the chance that the found mappings are indeed correct.

As we are interested in the leakage of topology information due to the overlay's routing algorithm, we focus on \emph{internal} attackers, where the adversary participates in a F2F overlay with one or more nodes $M \subset V$ under their control, which we call \emph{malicious nodes} in the following.
Protection against external attackers that infer overlay participants and links via traffic analysis is an orthogonal problem which can be addressed by tunneling F2F overlay messages through non-suspicious services~\cite{barradas2020towards}.

We assume that the attacker was able to identify a subset of the overlay's participants and lured each of them to let their node set up a link to at least one malicious node.
Malicious nodes participate in the embedding and routing but may deviate arbitrarily from correct behavior to obtain topology data.
In the following, we denote nodes of identified participants that are connected to malicious nodes as \emph{compromised nodes}.

\section{Inference of Topology Structure}
\label{sec:leakage}
As described in Section~\ref{sec:model}, we consider routing based on logical coordinates that are assigned to nodes based on a rooted spanning tree over the overlay network.
Because each link in the spanning tree corresponds to a unique link in the overlay network, it is desirable for the attacker to uncover the structure of the spanning tree, as it inherently corresponds to a subgraph of the overlay network's topology.
As the logical coordinate assigned to each node $u$ encodes the unique path in the spanning tree from $u$ to the root node, an attacker can leverage observations about which logical coordinates have been assigned to nodes to draw conclusions about the structure of the spanning tree and hence, the overlay.

To enable routing, data packets furthermore need to carry the logical coordinate of the recipient node.
As explained in the previous section, messages may be routed via shortcut links, i.e., links that are not part of the spanning tree.
By keeping track of which messages with which recipient coordinates have been routed via their nodes, the attacker can detect if a shortcut link has been used and infer possible paths taken by the message.
As a consequence, the actual routing of messages allows the attacker to make inferences with regards to shortcut links between nodes.  

In this section, we investigate the above risks in detail. 
To do so, we first formalize the concept of topological knowledge about an overlay network.
We then we specify which concrete inferences can be made from observed logical coordinates.
Afterwards, we analyze which inferences can be made from observations about the trajectories of messages routed via the overlay.

\subsection{Modeling topological knowledge}
For a given overlay network $O = (V,E)$, we model the adversaries' knowledge about $O$ at a fixed point in time by a tuple $(V_{obs},E_{obs},\overline{E},c_{obs})$.
$V_{obs}$ is a set of nodes that the adversary considers to be participating in the overlay.
This set always contains the compromised nodes defined in Section~\ref{sec:model} and the malicious nodes $M$ but may furthermore contain \emph{pseudonymous nodes} that the adversary is aware of but cannot immediately identify due to a lack of further information.
While the adversary can unambiguously relate malicious and compromised nodes to overlay nodes (e.g., by IP address), a pseudonymous node is considered to be participating in the overlay, but cannot be related to a particular overlay node.
More formally, the underlying injective mapping $\sigma : V_{obs} \rightarrow V$ is known to the adversary for malicious and compromised nodes but not for pseudonymous nodes.

$E_{obs}$ denotes links between nodes in $V_{obs}$ that the adversary knows to exist.
This means that it is guaranteed that if $(u,v) \in E_{obs}$, then $(\sigma(u),\sigma(v)) \in E$ holds.
The set $\overline{E_{obs}}$ encodes those links that the adversary knows to be non-existent between the nodes in $V_{obs}$, meaning that if $(u,v) \in \overline{E_{obs}}$ then $(\sigma(u),\sigma(v)) \notin E$.

The partial function $c_{obs} : V_{obs} \rightarrow \mathbb{N}_0^*$ encodes the assignment of logical coordinates of the nodes the adversary is aware of.
$c_{obs}$ is a partial function because the embedding algorithm may not yet have assigned a coordinate to a malicious or compromised node.
As we derive pseudonymous nodes from logical coordinates in the following, $c_{obs}$ is always defined for pseudonymous nodes. 

\subsection{Inference of tree links}
\label{sec:leakage-embedding}
We now consider concrete inferences made from observations about coordinates assigned to nodes.
A malicious participant may learn about the coordinates of other nodes in two ways:
\begin{itemize}
   \item To enable routing, each node needs to be aware of the logical coordinates of its neighbors. Therefore, as soon as a logical coordinate has been assigned to a node, it notifies all of its overlay networks about it. As a consequence, malicious nodes learn about the logical coordinates of their non-malicious neighbors.
   \item Messages carry the logical coordinate of the target node. If a message is routed via a malicious node, it can thus read the coordinate included in the message. 
\end{itemize}

\noindent Now consider that an adversary with knowledge $(V_{obs},E_{obs},c_{obs})$ has received a coordinate $c = (n_1,n_2,\ldots,n_l)$, with $l \geq 0$ and $\forall i \in \{1,\ldots,l\}: n_i \in \mathbb{N}_0$ that was previously unknown, meaning that there is no $u \in V_{obs}$ such that $c_{obs}(u) = c$.
First, they can obviously first infer that here exists a node $u$ to which coordinate $c$ has been assigned.
If $u$ is a compromised node, i.e., a non-malicious node with a malicious neighbor, then it is already included in $V_{obs}$ and only the mapping $c_{obs}(u) = c$ is added to $c_{obs}$.
Otherwise, the attacker generates a unique pseudonymous identifier for $u$, adds it to $V_{obs}$ and adds a corresponding mapping to $c_{obs}$.

\begin{figure}
  \centering
  \subfloat[][]{
     \def\svgwidth{0.475\columnwidth}
     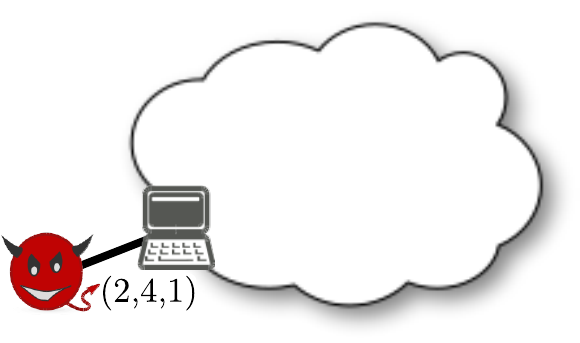
  }
  \hspace{0.5em}
  \subfloat[][]{
     \def\svgwidth{0.4\columnwidth}
     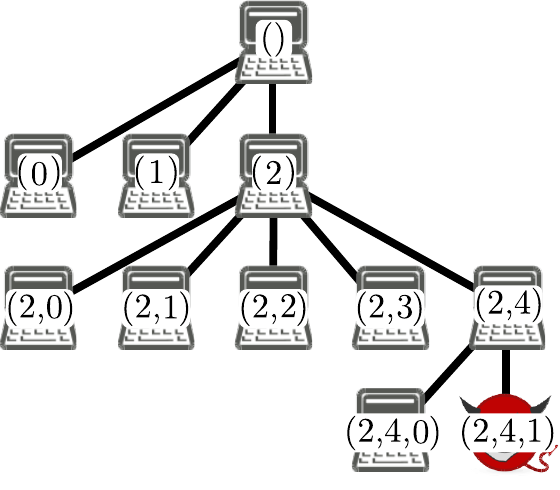
  }
  \caption{Example for inference of network structure from an observed coordinate. (a) A compromised node assigns coordinate $(2,4,1)$ to a malicious node. (b) From the received coordinate, the adversary infers the existence of pseudonymous nodes with coordinates $(2,4)$, $(2)$, and $()$ as well as tree links between them. For each of these coordinates, the attacker furthermore infers that nodes with lower coordinate elements, such as $(2,4,0)$, $(2,3)$, and $(1)$, exist.}
  \label{fig:example-inference}
\end{figure}

Furthermore, the attacker participates in the overlay and is thus aware of the embedding algorithm described in Section~\ref{sec:network-model}.
From coordinate $c$, they can thus also draw the following conclusions:
\begin{itemize}
   \item The coordinate assigned to a node corresponds to the coordinate of its parent node and an additional element at the end. Thus, if $l > 0$, meaning that $u$ is not the root node of the spanning tree, the they can infer that there must be a node $v$ with coordinate $c_v = (n_1,\ldots,n_{l-1})$ and that $u$ and $v$ are connected with each other. Thus, if $v$ is a previously unknown node, the attacker generates a unique identifier $ID_v$ for $v$, adds it to $V_{obs}$ and adds a corresponding mapping $c(ID_v) = (n_1,\ldots,n_{l-1})$. Furthermore, the attacker adds a link $(ID_v,u)$ to $E_{obs})$.
   \item The coordinate elements are determined by enumeration of child nodes in the spanning tree. Thus, if $n_l > 0$, they can infer that there must be nodes $v_0,v_1,\ldots,v_{n_l-1}$ with coordinates $(n_1,n_2,\ldots,n_{l-1},n_j)$ for $j \in \{0,\ldots,n_l - 1\}$ and that all of them are connected to the node with coordinate $c_v = (n_1,\ldots,n_{l-1})$. For those nodes whose coordinates were previously unknown, the attacker thus analogously generates unique identifiers and adds corresponding entries to $V_{obs}$, $c_{obs}$, and $E_{obs}$.
\end{itemize}
If $l > 1$, i.e., $u$ is not a child of the root node, the attacker can then additionally make analogous inferences for every non-empty prefix of $c$.
Figure~\ref{fig:example-inference} shows an example for inferences made from a coordinate based on the previously described considerations.
In Section~\ref{sec:study}, we present results from a simulation study that shed light on the number of nodes whose participation can be inferred in realistic settings.

\paragraph{Coordinate obfuscation}
To enable routing in a manner that hides the ultimate recipient of a message, Roos et al.~\cite{roos2016anonymous} proposed an obfuscation scheme for logical coordinates.
While not explicitly designed to hinder inference of topology structure, their obfuscation scheme nonetheless reduces the topological information an attacker can derive from observed coordinates.
In the following, we thus explain key changes and their effects in more detail.
In Section~\ref{sec:study}, we present simulation results showing that the obfuscation scheme drastically reduces the number of inferred participants.

Concretely, the embedding-based routing from Roos et al. differs from the routing presented in Section~\ref{sec:model} in two key points:
\begin{enumerate}
   \item Randomly chosen $b$-bit integers are used as coordinate elements instead of enumeration indexes.
   \item Before publishing the logical coordinate vector $c=(n_1,\ldots,n_l)$ of a node $u$, $c$ is padded to a fixed length by appending a corresponding number of additional randomly chosen integers. Subsequently, each element $n_i$ of the padded vector is replaced by a cryptographic hash value over $n_i$ and a randomly chosen number.
\end{enumerate}
Note that the second modification is used only to generate obfuscated addresses that can be published out of band to enable participants to contact a node in a privacy-preserving manner.
In the coordinate assignment procedure, nodes only use non-padded coordinates.

As a consequence of the first modification, an attacker learning about a coordinate $(n_1,n_2,\ldots,n_{l-1},n_l)$ cannot infer whether a coordinate $(n_1,n_2,\ldots,n_{l-1},n')$ with $n' \neq n_l$ has been assigned to any other node, since $n_l$ was chosen randomly, independent of the number of children in the spanning tree.
However, the attacker can still infer that there is another node to which the coordinate $(n_1,n_2,\ldots,n_{l-1})$ has been assigned and that this node is connected to the node with coordinate $(n_1,\ldots,n_l)$ and they can proceed analogously with every non-empty prefix of the coordinate.

The second modification keeps the attacker from learning about previously unknown node coordinates by reading the target coordinates included in data messages routed via malicious nodes.
Shortly, this is because the attacker cannot determine the actual number of randomly added elements of the target coordinates.
While the attacker, given an obfuscated target coordinate $\hat{c}=(n_1,n_2,\ldots,n_l)$, can determine the longest common prefix $n_1,\ldots,n_k$ between $\hat{c}$ and any coordinate they are already aware of, they cannot tell whether element $n_{k+1}$ of $\hat{c}$ is already part of the random padding or not.
Due to the properties of the cryptographic hash function, the attacker furthermore cannot unambiguously infer the value of the $k+1$-th element of the padded coordinate.
Even if a node $u$ publishes multiple obfuscated variants of its coordinate, the attacker can only determine possible longer common prefixes among them by exhaustive search over the range of possible element values, which is computationally infeasible for a sufficiently large value of $b$.

\begin{figure}
  \centering
  \def\svgwidth{0.55\columnwidth}
  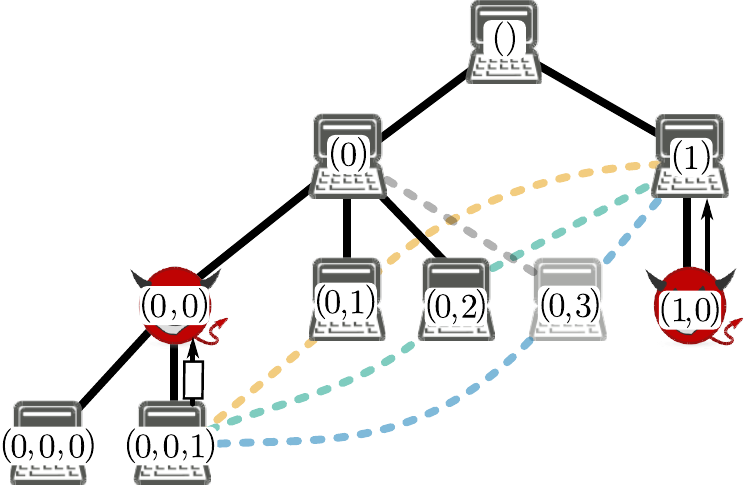
  \caption{Example for an ambiguous route. Solid lines denote known tree links and dashed lines denote unknown but possible links. The different link colors highlight possible message trajectories. The malicious node $(1,0)$ first sent a message $m$ with target $(0,0,0)$ to its parent. When malicious node $(0,0)$ afterwards receives $m$ from the compromised child node $(0,0,1)$, $m$ may have been routed either through node $(0,1)$, node $(0,2)$ or a yet unknown node $(0,3)$.}
  \label{fig:example-challenges}
\end{figure}

\subsection{Inference of shortcut links}
\label{sec:leakage-messages}
Recall from Section~\ref{sec:model} that embedding-based routing also considers non-tree edges for forwarding.
To detect the usage of such shortcut links, malicious nodes record every message $m$ that they received, including the message's target coordinate $c_t$ as well as the coordinate of the neighbor $c_n$ from which they received the message.
If the attacker is aware of the logical coordinate $c_p$ of another node over which the message was routed previously, they can then check if $c_n$ is a prefix of $c_p$ or of $c_t$.
If this is not the case, then $c_n$ does not lie on the path from $c_p$ to $c_t$ in the spanning tree underlying the coordinate assignment and thus, $m$ must have been routed via a shortcut link.

An attacker may become aware of the coordinates of nodes previously traversed by a message via multiple means.
If the originator $u$ of a message $m$ writes its own logical coordinate $c_u$ into $m$ to enable the recipient of $m$ to send a reply, the attacker can simply read the value of $c_u$.
In the following, we however do not assume that sending nodes include their coordinate in messages sent.
Even if they do so, reading the sender coordinate by malicious nodes can be prevented by having nodes publish a cryptographic key along with their coordinate, such that senders can attach their coordinate to messages in an encrypted form.
Since F2F networks typically do not obfuscate message contents during routing, e.g., via re-encryption, the adversary can instead determine if the same message was routed via two or more malicious nodes and in which order.

Given the adversary has received a message and is aware of the coordinate of a previously traversed node, the actual inference of possible shortcut links is non-trivial.
The message may have been routed via a yet unknown node or there may be two or more known nodes that qualify as the next hop.
Figure~\ref{fig:example-challenges} shows an example for such cases.

To formalize the conditions when the existence or absence of a link can be concluded, we first introduce the concept of a \emph{hypothetical overlay} that addresses the possible presence of yet unknown nodes.
Afterwards, we define the notion of a \emph{plausible trajectory} within a hypothetical overlay and subsequently specify when a message is said to \emph{prove} the existence or absence of an overlay link.

\subsubsection{Hypothetical overlay}
Given knowledge $(V_{obs},E_{obs},\overline{E_{obs}},c_{obs})$ about an overlay, a corresponding \emph{hypothetical overlay} is a tuple $(V_{obs}, V_D, E_H, c_H)$, where $V_D$ denotes a set of \emph{dummy nodes}, $E_H \subseteq (V_{obs} \cup V_D) \times (V_{obs} \cup V_D)$ and $c_H : (V_{obs} \cup V_D) \rightarrow \{0,\ldots,2^b-1\}^*$.

\begin{figure}
  \centering
  \subfloat[][]{
     \def\svgwidth{0.15\textwidth}
     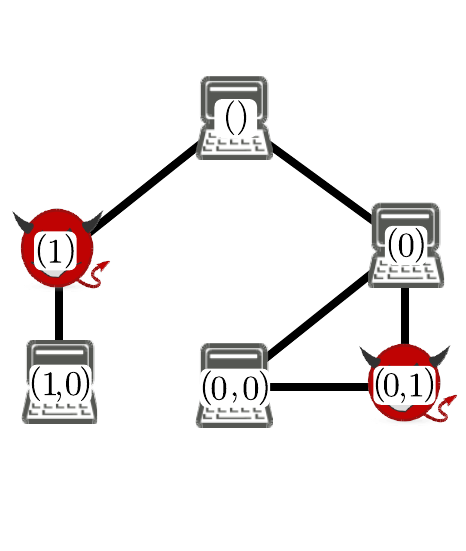
  }
  \hspace{1em}
  \subfloat[][]{
     \def\svgwidth{0.25\textwidth}
     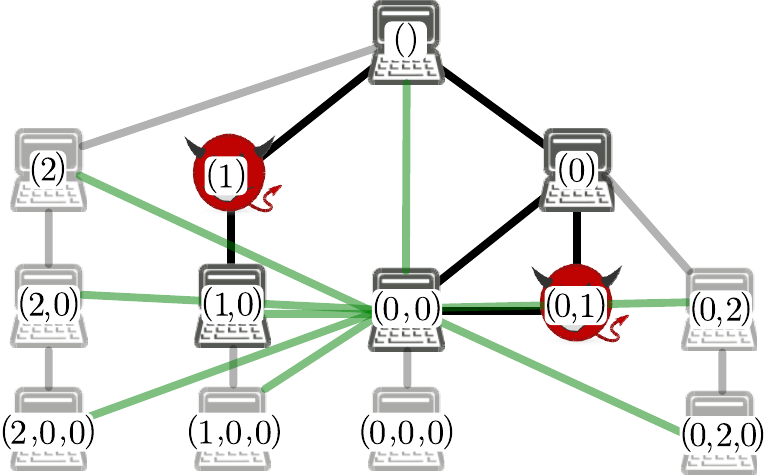
  }
  \caption{Example for generation of hypothetical overlay from adversary knowledge. (a) Derivation of initial overlay from a priori knowledge. (b) Introduction of dummy nodes that represent unknown nodes as well as possible links between all nodes. For better readability, only the hypothetical links starting from node $(0,0)$ are displayed as green lines.}
  \label{fig:hypo-generation}
\end{figure}

Each dummy node in $V_D$ represents an \emph{unknown number of nodes} with the same parent in the spanning tree.
The coordinate assignment $c_H$ assigns the same coordinates to each node from $V_{obs}$ as $c_{obs}$ but additionally assigns a unique, random coordinate to every dummy node.
To enable discovery of all possible trajectories, $E_H$ includes all pairs of nodes $(u,v) \in {(V_{obs} \cup V_D)}^2, u \neq v$ except those for which $(u,v) \in \overline{E_{obs}}$.
Given knowledge $(V_{obs},E_{obs},\overline{E_{obs}},c_{obs})$ and a set of malicious nodes $M$, a corresponding hypothetical overlay can be generated via the following steps:
\begin{enumerate}
   \item Set $V_D = \emptyset$, $E_H = E_{obs}$, and $c_H = c_{obs}$ 
   \item Determine the length $l_{max}$ of the longest coordinate in $c_{obs}$
   \item For every non-malicious node in $V_{obs}$ with coordinate length $l \leq l_{max}$, add a subtree of depth $l_{max} - l + 1$ by adding a dummy node to $V_D$ with a unique coordinate for each level.
   \item For every pair of nodes $u,v \in (V_{obs} \cup V_D) \setminus M$ with $(u,v) \notin \overline{E_{obs}}$, add a link $(u,v)$ to $E_H$.
\end{enumerate}
Figure~\ref{fig:hypo-generation} shows an example for the generation of the hypothetical overlay.

While a node may have a shortcut link to a yet unknown node whose logical coordinate has more than $l_{max}+1$ elements, we omit the generation of such dummy nodes.
It can easily be shown that if a message may have been routed via an unknown node $u$ with a longer coordinate, then it is also possible that this message was routed instead via the predecessor of $u$ whose coordinate has length $l_{max}+1$, which is represented by a dummy node.
Thus, even if dummy nodes with coordinates longer than $l_{max}+1$ elements are omitted from the hypothetical overlay, we ensure that if a message may have been routed via an unknown node, then there always is at least one corresponding route via a dummy node in the hypothetical overlay.

\subsubsection{Plausible trajectories and link existence}
To be able to define a plausible trajectory, we first need to formalize the observation of a message by malicious nodes.
We do so with the notion of a \emph{trace record}, as given by Definition~\ref{def:trace}.

\begin{definition}
   \emph{(Trace record)}
Let $O = (V,E)$ be an overlay network and let $M \subset V$ be a set of observation points in $O$.
   For a message $m$, let $p = u_1,u_2,..,u_k$ with $k \in \mathbb{N}$ and $\forall i \in \{1,..,k\}: u_i \in V$ be the path along which $m$ has been forwarded in $O$.

   For a given pair $m_b, m_e \in M$, a \emph{trace record} of $m$ on $p$ is a 4-tuple $(m_s,u_s,u_e,m_e)$ where
   \begin{enumerate}
      \item $u_s,u_e \in V \setminus M$
      \item There exists a subsequence $v_1,v_2,..,v_l$ of $p$ such that $v_1 = m_s, v_2 = u_s, v_{l-1} = u_e, v_l = m_e$ and $\forall i \in \{3,..,l-2\}: v_i \notin \{u_s,u_e\}$.
   \end{enumerate}
   \label{def:trace}
\end{definition}

Although a packet may traverse more than two malicious nodes on its way to the target node, we treat each path between two consecutively traversed malicious nodes as a separate trace record.
We consider this simplification to be valid, as the greedy routing of each message from a malicious node $m$ to another $m'$ is independent from the path over which the message was routed to $m$ before.

Based on the notion of a trace record, we define a \emph{plausible trajectory} as given by Definition~\ref{def:plausible-trajectory}.
\begin{definition}
   \emph{(Plausible trajectory)}
   Let $O = (V,E)$ be an overlay network and let $c : V \rightarrow \mathbb{N}_0^*$ be a coordinate assignment to the nodes in $O$. 
   Furthermore, let $M \subset V$ denote a set of observation points, $K = (V_{obs},E_{obs},\overline{E_{obs}}, c_{obs})$ denote a priori knowledge about $V$ and $E$, and let $(V_{obs},V_D,E_H,c_H)$ be a hypothetical overlay for $K$.

   Given a trace record $r = (m_s,u_s,u_e,m_e)$ with $m_s,m_e \in M$ and $u_s,u_e \in V_{obs}$ of a message with target coordinate $c_t$, a sequence of nodes $v_1,v_2,..,v_l$ from $V_{H}$ is called a \emph{plausible trajectory} for $r$ towards $c_t$ given knowledge $K$ if:
   \begin{enumerate}
      \item $v_1=m_s$,$v_2=u_s$,$v_{l-1}=u_e$,$v_l=m_e$
      \item $\forall i \in \{2,..,l-1\}: v_i \notin M$
      \item $\forall i \in \{1,..,l-1\}: (v_i,v_{i+1}) \in E_H$
      \item $\forall i \in \{1,..,l-1\}: \delta(c(v_{i+1}),c_t) < \delta(c(v_{i}),c_t)$
      \item $\forall i \in \{1,..,l-1\}: $ \[ \delta(c(v_{i+1}),c_t) \leq \displaystyle\min_{\{u \in V_{obs} \mid (v_i,u) \in E_{obs}\}} \delta(c(u),c_t) \]
   \end{enumerate}
   \label{def:plausible-trajectory}
\end{definition}
The first condition of Definition~\ref{def:plausible-trajectory} ensures that only trajectories matching the trace record are considered to be plausible.
Because we only consider trajectories between two malicious nodes, the second condition ensures that other malicious nodes are excluded.
The third condition ensures that a plausible trajectory does not contradict the adversaries' knowledge about absent links, as pairs $(u,v) \in \overline{E_{obs}}$ are not included in $E_H$.
The fourth condition reflects that, due to greedy routing, nodes only forward messages to neighbors whose distance to the target is strictly lower than their own.
The fifth condition furthermore guarantees that a plausible trajectory does not contradict the adversaries' knowledge about existing links.
For an example, again consider Figure~\ref{fig:example-challenges}.
If there would be a link between node $(0,1)$ and $(0,0,0)$ that is known to the adversary, then the fifth condition would ensure that any route via $(0,1)$ is not considered plausible. 
Because if the message for target $(0,0,0)$ had been received by node $(0,1)$, then it would have greedily forwarded it directly to $(0,0,0)$ instead of forwarding it to node $(0,0,1)$.

Although there may be multiple plausible trajectories for a given trace record, there are cases where the adversary may nonetheless be able to infer the existence or absence of a link.
Definition~\ref{def:proof-existence} therefore specifies when a trace record is said to \emph{prove} the existence or absence of a link between known nodes.

\begin{definition}
   \emph{(Proof of link existence)}
   Let $O = (V,E)$ be an overlay network and let $c : V \rightarrow \mathbb{N}_0^*$ be a coordinate assignment to the nodes in $O$. 
   Furthermore, let $M \subset V$ denote a set of observation points, $K = (V_{obs},E_{obs},\overline{E_{obs}}, c_{obs})$ denote a priori knowledge about $V$ and $E$.

   A trace record $r = (m_s,u_s,u_e,m_e)$ with $m_s,m_e \in M$ and $u_s,u_e \in V_{obs}$ of a message with target coordinate $c_t$ \emph{proves the existence of a link between two known nodes $u,v \in V_{obs} \setminus M$ given knowledge $K$}, if all plausible trajectories for $r$ towards $c_t$ given knowledge $K$ include the sequence $u,v$.

   \label{def:proof-existence}
\end{definition}

\subsubsection{Limits of inference from data messages}
For an attacker aiming to perform graph-based de-anonymization attacks, it is desirable to obtain knowledge about the links incident to pseudonymous nodes, as this can be used to make correct de-anonymization more likely.
In the following we show that by tracing message trajectories, the adversary cannot unambiguously infer shortcut links between compromised nodes.
In particular, we show that for every pair $(u,v)$ of nodes where $u$ or $v$ is a pseudonymous node, every trace record that has a plausible trajectory that includes the sequence $u,v$ also has at least one plausible trajectory that does not include the sequence $u,v$.

As a prerequisite, Lemma~\ref{lem:tree-dst} states that whenever the logical coordinates of two nodes $u$ and $v$ have the same length, then either $\delta(c(u),t) = \delta(c(v),t)$ or $\delta(c(u),t) \geq \delta(c(v),t) + 2$ for every coordinate $t \in \mathbb{N}_0*$.
\begin{lemma}
   Let $O = (V,E)$ be an overlay network and let $c : V \rightarrow \mathbb{N}_0^*$ be a coordinate assignment for the nodes in $V$.
   The following holds for every pair of nodes $u,v \in V$: if $|c(u)| = |c(v)|$, then there is no coordinate $t \in \mathbb{N}_0^*$ such that $\delta(c(u),t) = \delta(c(v),t) + 1$.
   \label{lem:tree-dst}
\end{lemma}
\begin{proof}
   As described in Section~\ref{sec:model}, the tree distance between two coordinates is computed solely from the length of the coordinates as well as the length of their common prefix.
   Since $|c(u)| = |c(v)|$, $v$ can only have a lower distance to $t$ if it has a longer common prefix.
   Therefore, assume that $cpl(c(v),t) = cpl(c(u),t)+k$ with $k \in \mathbb{N}, k \geq 1$. Thus,
   \begin{align*}
      \delta(c(v),t) &= |c(v)| + |c(t)| - 2 cpl(c(v),t) \\
      &= |c(v)| + |c(t)| - 2 cpl(c(u),t) - 2k \\
      &= |c(u)| + |c(t)| - 2 cpl(c(u),t) - 2k \\
      &= \delta(c(u),t) - 2k \qedhere
  \end{align*}
\end{proof}
Using Lemma~\ref{lem:tree-dst}, we can now prove Theorem~\ref{thm:compromised-only}.
\begin{theorem}
   Let $O = (V,E)$ be an overlay network and let $c : V \rightarrow \mathbb{N}_0^*$ be a coordinate assignment to the nodes in $O$. 
   Also, let $M \subset V$ denote a set of observation points, $K = (V_{obs},E_{obs},\overline{E_{obs}}, c_{obs})$ denote a priori knowledge about $V$ and $E$.
   Furthermore, let $u,v \in V_{obs} \setminus M$ be a pair of nodes such that $(u,v) \notin E_{obs} \cup \overline{E_{obs}}$.

   If there is a trace record $r = (m_s, u_s, u_e, m_e)$ with $m_s,m_e \in M$ and $u_s,u_e \in V_{obs}$ of a message with a target coordinate $c_t$ that proves the existence of a link between $u$ and $v$, then it must hold that $u=u_s$ and $v=u_e$.
   \label{thm:compromised-only}
\end{theorem}
\begin{proof}
   In the following, we show that if $v$ does not have a malicious neighbor, the adversary cannot unambiguously determine whether any message was indeed forwarded directly from $u$ to $v$ or vice versa.
   This is because it is always possible that there is a yet unknown node over which $u$ or $v$ may have routed the message instead.
   More formally, we show that if $v$ is not a compromised node, i.e. , not connected to a malicious node, then for every trace record, there is at least one plausible trajectory that does not include the sequence $u,v$ or $v,u$. 
   Thus, no trace record proves the existence of a link $(u,v)$ according to Definition~\ref{def:proof-existence}. The proof analogously holds for the case that $u$ is not a compromised node.

   W.l.o.g., assume that $v$ is not a compromised node.
   First, note that $u$ cannot be a child of $v$ and vice versa.
   Otherwise, since the attacker is aware of $u$'s and $v$'s coordinates, he could already infer the existence of a link between $u$ and $v$ as described in Section~\ref{sec:leakage-embedding}.
   Nonetheless, it is possible that $u$ is a higher order descendant of $v$ in the sense that $u$ may be a descendant of a child of $v$ and vice versa.
   In the following, we first present the proof for the case that neither $u$ is a descendant of $v$ nor $v$ a descendant of $u$ and subsequently proof the case where one is a descendant of the other.

   Given that $u$ is not a descendant of $v$ and vice versa, it follows that neither of them can be the root node of the spanning tree.
   As all neighbors of $v$ are non-malicious by assumption, $v$ thus must have a parent $p_v$ that also must be a non-malicious node.
   $v$ also must have at least one more non-malicious neighbor the adversary is aware of, which may either be a child of $v$ or a neighbor connected via an already known shortcut link.
   Otherwise, the adversary is unable to tell if it is even possible that any message he received traversed $v$.

   The two key insights used in this proof are that the adversary cannot tell i) if there is another, unknown child of $p_v$ besides $v$, and ii) if $v$ has any yet unknown children.
   Thus, the hypothetical overlay $(V_{obs},V_D,E_H,c_H)$ corresponding to the adversaries knowledge $K$ contains a dummy node $sib_v$ that is a sibling of $v$ as well as a dummy node $chd_v$ that is a child of $v$.
   As the attacker is unaware of the connections of these unknown nodes, $sib_v$ is connected to $u$ as well as all neighbors of $v$, as it is possible that an unknown child of $p_v$ may have such links.
   Similarly, $chd_v$ is also connected to $u$.
   We consider a worst case scenario, where $u$ is neither connected to $p_v$ nor $n_v$ and the attacker is aware of this fact, such that $u$ is also neither connected to $p_v$ nor $n_v$ in the hypothetical overlay.
   Figure~\ref{fig:proof-hypo-model} illustrates the considered scenario.
\begin{figure}
  \centering
     \def\svgwidth{0.25\textwidth}
     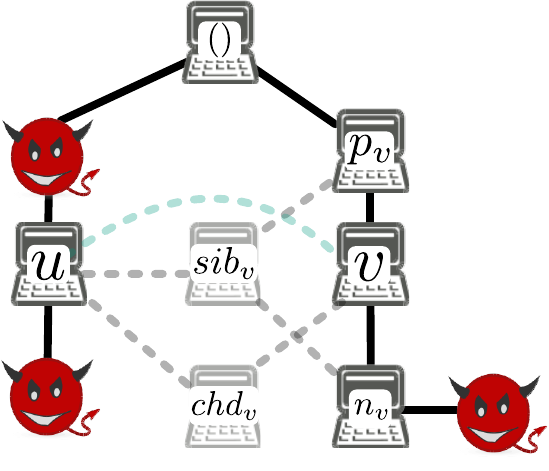
     \caption{Hypothetical overlay considered for the proof of Theorem 1. Solid lines denote known tree links and dashed lines denote possible links added to the hypothetical overlay. The blue dashed line is the one whose existence attacker tries to infer. $p_v$ and the parent of $u$ must not be a child of the root node but can be any descendant. Also $n_v$ does not have not be the child of $v$ but can be a neighbor connected via a known shortcut link.}
  \label{fig:proof-hypo-model}
\end{figure}
   A message with target $c_t$ may only be forwarded from $u$ to $v$ or vice versa if $\delta(c_{H}(u),c_t) < \delta(c_{H}(v),c_t))$ or $\delta(c_{H}(u),c_t) > \delta(c_{H}(v),c_t))$, respectively.
   We now proof each case separately.

   \textbf{Case $\delta(c_{H}(u),c_t) < \delta(c_{H}(v),c_t)$:} In this case, the message with target $c_t$ forwarded by a malicious node must subsequently have been routed first via $v$ and afterwards via $u$.
   Since $u$ is not a descendant of $v$, in this case it follows that the coordinate $c_H(v)$ of $v$ must not be a prefix of $c_t$, as then only descendants of $v$ can have a lower distance to $c_t$ than $v$.
   Because all neighbors of $v$ are non-malicious, any message forwarded by a malicious node towards $v$ must first traverse one of $v$'s neighbors $n_v$ before reaching $v$.
   Since $v$'s coordinate is not a prefix of $c_t$, $n_v$ cannot be $v$'s parent $p_v$, because it must hold that $\delta(c_{H}(p_v),c_t) < \delta(c_{H}(v),c_t))$, meaning that $p_v$ would not forward the message to $v$. Thus, $n_v$ must either be a child of $v$ or a neighbor connected via a known shortcut link.

At the same time, $n_v$ may also be connected to another, unknown sibling of $v$, and thus $n_v$ is connected to $sib_v$ in the hypothetical overlay.
Because $v$ and $sib_v$ are a child of $p_v$, it must hold that $|c_H(v)| = |c_H(v')|$. 
Also, since $c_H(v)$ is not a prefix of $c_t$, it follows that $cpl(c_H(sib_v),c_t) \geq cpl(c_H(v),c_t)$ and thus, $\delta(c_H(sib_v),c_t) \leq \delta(c_H(v),c_t)$.
As a consequence, $n_v$ may thus send the message to $sib_v$ instead of $v$.
Since $sib_v$ is connected to $u$ in the hypothetical overlay, there is at least one plausible trajectory that includes the sequence $sib_v,u$ instead of $v,u$ and therefore, any trace record obtained from such a message cannot prove the existence of the link $(u,v)$.

\textbf{Case $\delta(c_{H}(u),c_t) > \delta(c_{H}(v),c_t)$:} In this case, a message was first sent to $u$, which then may have forwarded it to $v$.
Here, we distinguish between three cases, namely that i) $|c_{H}(u)| = |c_{H}(v)|$,  ii) $|c_{H}(u)| < |c_{H}(v)|$, and iii) $|c_{H}(u)| > |c_{H}(v)|$.

In case (i), Lemma~\ref{lem:tree-dst} implies that $\delta(c_{H}(u), c_t) \geq \delta(c_{H}(v),c_t)) + 2$.
For every child $d$ of $v$, it therefore holds that $\delta(c_{H}(u), c_t) \geq \delta(c_{H}(v),c_t)) + 1$.
Consequently, there is at least one plausible trajectory that includes the sequence $u,chd_v$ instead of $u,v$ and thus, any trace record obtained in this case cannot prove the existence of link $(u,v)$.

In case (ii), the assumption $\delta(c_{H}(u),c_t) > \delta(c_{H}(v),c_t)$ implies that the coordinate $c_{H}(p_v)$ of $v$'s parent must be a prefix of $c_t$ and thus $\delta(c_{H}(u),c_t) > \delta(c_{H}(p_v),c_t)$. 
From Lemma~\ref{lem:tree-dst}, it follows that $\delta(c_{H}(u), c_t) \geq \delta(c_{H}(p_v),c_t)) + 2$ and thus $\delta(c_{H}(u), c_t) \geq \delta(c_{H}(d),c_t)) + 1$ for every child $d$ of $p_v$, including those represented by $sib_v$ in the hypothetical overlay. 
Thus, there is at least one plausible trajectory that includes the sequence $u,sib_v$ instead of $u,v$, such that any trace record obtained in this case also cannot prove the existence of link $(u,v)$.

In case (iii), we further need to distinguish two cases, namely that a) $cpl(c_H(v),c_t) < |c_H(v)|$, i.e., the recipient of the message is not a descendant of $v$, and b) $cpl(c_H(v),c_t) = |c_H(v)|$, i.e., the recipient is a descendant of $v$.
If (a) holds, then it is possible that $p_v$ has another child represented by $sib_v$ with $cpl(c_H(sib_v),c_t) = cpl(c_H(v),c_t)$ and thus $\delta(c_H(sib_v),c_t) = \delta(c_H(v),c_t)$.
As it then follows that $\delta(c_H(u),c_t) > \delta(c_H(sib_v),c_t)$, it is thus possible that $u$ sent the message to $sib_v$ instead of $v$.
If (b) is true, it must hold that $cpl(c_H(u),c_t) \leq cpl(c_H(p_v),c_t) = cpl(c_H(v),c_t)-1$, since $u$ is not a descendant of $v$.
Thus,
\begin{align*}
   \delta(c_{H}(u),c_t) &= |c_{H}(u)|+|c_t|-2cpl(c_{H}(u),c_t) \\
                        &\geq |c_{H}(v)| + 1 +|c_t|-2cpl(c_{H}(p_v),c_t) + 2 \\
                        &= \delta(c_{H}(p_v),c_t) + 3
   \end{align*}
Thus, for every child $d$ of $v$, it holds that $\delta(c_H(u),c_t) > \delta(c_H(d),c_t) + 2$.
As a consequence, there is at least one plausible trajectory that contains the sequence $u,chd_v$ and therefore any trace record obtained in this setting also cannot prove the existence of link $(u,v)$.

  We now consider the case that either $u$ is a descendant of $v$ or vice versa.
  As explained at the beginning of the proof, $u$ cannot be a child of $v$ or vice versa.
  Furthermore, there must be at least one malicious node on the path between $u$ and $v$ in the spanning tree.
  Otherwise, it is always possible that the message was routed solely along the tree links.

  \textbf{Case $\delta(c_{H}(u),c_t) < \delta(c_{H}(v),c_t)$:}
  In this case, the message may have first been routed via $v$, which then may have sent it to $u$. 
  Since $v$ is not a neighbor of a malicious node, any message that may traverse $v$ must first traverse a non-malicious neighbor $n_v$ of $v$. 
  Furthermore, it must hold that $\delta(c_{H}(v),c_t) < \delta(c_{H}(n_v),c_t)$, as $n_v$ would otherwise not forward the message to $v$.

  First, consider the case that $v$ is descendant of $u$.
  Since $u$ cannot be the parent of $v$, it must also hold that $p_v$ is a descendant of $u$ and $\delta(c_{H}(u),c_t) < \delta(c_{H}(p_v),c_t)$.
  From latter statement, it also follows that the target of the message cannot be a descendant of $p_v$ and thus, it must hold that $\delta(c_{H}(w),c_t) = \delta(c_{H}(p_v),c_t) + 1$ for every child $w$ of $p_v$.

  Because all children of $p_v$ have the same distance to $c_t$, $v$'s neighbor $n_v$ may thus forward the message to $v$'s sibling $sib_v$ instead of $v$.
  As $sib_v$ is connected to $u$ in the hypothetical overlay, there is at least one plausible trajectory that includes the sequence $sib_v,u$ instead of $v, u$ and therefore, any trace record obtained from such a message cannot prove the existence of the link $(u, v)$.

  Now consider the case that $u$ is a descendant of $v$.
  In this case, $v$ must have a child $w$ that also has $u$ as descendant.
  Furthermore, it must hold that $\delta(c_{H}(w),c_t) = \delta(c_{H}(v),c_t) - 1$.
  Because $u$ is also a descendant of $w$, $w$ also must have a child $w'$ such that $\delta(c_{H}(w'),c_t) = \delta(c_{H}(w),c_t) - 1 = \delta(c_{H}(v),c_t) - 2$.
  For every other child $y$ of $w$, it must hold that $\delta(c_{H}(y),c_t) = \delta(c_{H}(w),c_t) + 1 = \delta(c_{H}(v),c_t)$.
  As a consequence, it is possible that $n_v$ has forwarded the message to a yet unknown child represented by $chd_w$ of $w$ instead.
  Since $chd_w$ is connected to $u$ in the hypothetical overlay, there is at least one plausible trajectory that includes the sequence $chd_w,u$ instead of $v,u$.

  \textbf{Case $\delta(c_{H}(u),c_t) > \delta(c_{H}(v),c_t)$:}
  If $v$ is a descendant of $u$, then it must hold that $\delta(c_{H}(v),c_t) < \delta(c_{H}(u),c_t) - 1$, since $u$ is not the parent of $v$.
  As a consequence, for every child $w$ of $v$, it must hold that $\delta(c_{H}(w),c_t) < \delta(c_{H}(u),c_t)$.
  Thus, $u$ may forward the message to a yet unknown child of $v$ represented by $chd_v$ in the hypothetical overlay.
  Consequently, there is a plausible trajectory that includes the sequence $u,chd_v$ instead of $u,v$.

  If $u$ is a descendant of $v$, then it must hold that $\delta(c_{H}(u),c_t) < \delta(c_{H}(v),c_t) - 1$, since $v$ is not the parent of $u$.
  Analogously to the previous case, it thus follows that there is a plausible trajectory that includes the sequence $u,chd_v$ instead of $u,v$.
\end{proof}

Note that Theorem~\ref{thm:compromised-only} holds irrespective of whether the target coordinate of the message is obfuscated, as described in Section~\ref{sec:leakage-embedding}. 
However, if the target coordinates are not obfuscated, an adversary that inspects received messages may eventually learn almost all coordinates that are currently assigned to nodes and thus become more confident about the absence of yet unknown nodes.
For scenarios where the overhead incurred by coordinate obfuscation is considered too high, our proof of Theorem~\ref{thm:compromised-only} suggests that the deliberate introduction of fake children nodes by nodes that actually have only a single child node is a protection measure worth further investigation.

One limitation of the proof is that it is restricted to settings where the adversary cannot determine the coordinate of the actual originator of the message.
However, as explained before, the coordinate of the sender can be obfuscated via different means to prevent monitoring by traversed nodes.

\section{Simulation study}
\label{sec:study}
In the previous section, we showed that malicious participants can unambiguously infer tree links from observed coordinates while the monitoring of message trajectories does not allow unambiguous inferences most of the time.
While the obfuscation scheme from VOUTE~\cite{roos2016anonymous} outlined in Section~\ref{sec:leakage-embedding} can be used to render the coordinates included in data packets useless for inference of tree links, it does not prevent inferences from the coordinates propagated by the embedding algorithm.

To evaluate the privacy risk posed by the fact that every node learns the actual logical coordinate of each of its neighbors in realistic settings, we performed a simulation study using OMNet++.
In particular, this study investigates how many previously unknown nodes malicious participants can infer from the observed coordinates.
We chose to do a simulation instead of a measurement study on a real-world F2F overlay, since the routing considered in this paper is not yet in use by any currently deployed overlay.

\textbf{Metrics:} Given an adversary with malicious node set $M$ and knowledge $K=(V_{obs},E_{obs},\overline{E_{obs}}, c_{obs})$, let $O_{obs} = (V_{obs},E_{obs})$ denote the overlay the attacker is aware of after it has processed the logical coordinates assigned to the neighbors of malicious nodes.
We measure the number of newly discovered nodes by the number of pseudonyms $N_p = |\{ u \mid u \in V_{obs} \wedge u \notin N_{O_{obs}}(M) \}|$. 
As the adversary can only de-anonymize nodes he is aware of, $N_p$ thus gives the maximum number of users the adversary may de-anonymize based on routing information.

\textbf{Datasets:} Because existing F2F overlays have not yet reached widespread adoption and are designed to hinder collection of topology information, there are currently no network snapshots available for investigation.
Given that F2F overlays resemble social trust relationships, we thus leverage datasets obtained from crawling online social networks, whose characteristics are presented in Table~\ref{tab:datasets}.
All of these graphs are undirected.

\emph{SPI} denotes a graph obtained from a German university social network~\cite{paul2016students}.
Brightkite (BK) denotes a graph obtained by crawling the Brightkite location-based online social network~\cite{cho2011friendship}.
\emph{WoT} represents a subgraph of a snapshot from the PGP Web of Trust taken on February 7, 2012 from the \texttt{wotsap}-database\footnote{\url{www.lysator.liu.se/~jc/wotsap/wots2/}, accessed 2021-01-08}.
As the original snapshot was a directed graph, we first removed any links between pairs of nodes that do not have links in both directions.
The WoT graph used for our study consists of the largest connected component of the modified snapshot.

\begin{table}
\small
   \renewcommand{\arraystretch}{1.3}
   \setlength\tabcolsep{1mm}
   \caption{Number of nodes $n$, median degree $deg_{med}$, maximum degree $deg_{max}$, average shortest path length $\overline{spl}$ and clustering coefficient $cc$ of the graph datasets.}
   \label{tab:datasets}
   \centering
   \begin{tabular}{|c|c|c|c|c|c|}
      \hline
      \textbf {Graph} & \textbf{n} & \textbf{$deg_{med}$} & \textbf{$deg_{max}$} & \textbf{$\overline{spl}$} & $cc$ \\ \hline
      SPI & 9,222 & $7$ & $147$ & $4.67$ & $0.34$ \\ \hline 
Web-of-Trust & 37,937 & $2$ & $1,074$ & $6.28$ & $0.475$ \\ \hline 
Brightkite & 56,739 & $2$ & $1,134$ & $4.92$ & $0.268$ \\ \hline
Facebook & 63,392 & $11$ & $1,098$ & $4.3$ & $0.15$ \\ \hline
   \end{tabular}
\normalsize
\end{table}

\textbf{Model, System Parameters, and Set-up:} For our study, we implemented two state of the art embedding algorithms, namely Greedy Forest Routing (GFR)~\cite{houthooft2015robust} and VOUTE~\cite{roos2016anonymous}.
In contrast to GFR, which uses enumeration indexes as coordinate elements, VOUTE uses random numbers, thus preventing the inference of further sibling nodes and their coordinates.
While both algorithms allow the redundant construction of multiple embeddings, we chose the parameters of the algorithms such that a single BFS spanning tree with a randomly chosen root node is constructed in each simulation run.
Since each embedding assigns a different logical coordinate to every node, the inference of network structure across multiple parallel embeddings is non-trivial.
We thus consider this task to be an interesting venue for further research.

As our adversary can only obtain information from compromised nodes, we performed simulations with $N_C \in \{200, 400, 600, 800, 1000\}$ compromised nodes for each graph.
We use fixed values for $N_C$ instead of a fraction of the graph's number of nodes, as this allows us to focus on the effect of graph structure on the effectiveness of the attack.
Otherwise, we cannot tell if an increase in the number of inferred pseudonymous for large graphs stems mostly from the increased number of compromised nodes.

For each graph, we determined the compromised nodes by randomly selecting a subset of nodes from $G$ that serve as malicious nodes.
For each value of $N_C$, we randomly selected 20 sets of malicious nodes such that each set results in $N_C$ compromised nodes.

We implemented two types of adversarial behaviors:
In the first scenario, the malicious nodes follow the embedding algorithm correctly.
In the second scenario, each malicious node acts to each non-malicious neighbor $u$ as if it does not have other neighbors, thus always becoming a child node of $u$.
The only exception is that if a malicious node is chosen as root node, it follows the embedding algorithm correctly.
Whenever a compromised node $u$ becomes the child of a malicious node $v$, the coordinate of $u$ only reveals pseudonymous nodes that can already be inferred from $v$'s coordinate.
Thus, we expect the number of inferred pseudonyms $N_p$ to increase when the malicious nodes actively keep non-malicious nodes from becoming their child.

Each simulation run for a given graph $G=(V,E)$ and set $M \subset V$ of malicious nodes proceeded as follows: first, create a network with $|V|$ nodes and add a corresponding link for each edge in $E$.
Subsequently, configure the nodes in $M$ according to the simulated adversarial behavior and initialize the adversaries' knowledge $K=(V_{obs},E_{obs},\overline{E_{obs}},c_{obs})$ such that $V_{obs} = M \cup N_G(M)$, $E_{obs}=\{(u,v) \mid (u,v) \in E \wedge u,v \in M \cup N(M)\}, \overline{E_{obs}} = \emptyset$, and $c_{obs} = \emptyset$.
Afterwards, run the simulation until all nodes have received a coordinate. Whenever a compromised node notifies a malicious neighbor about its coordinate $c$, any tree links and coordinates that can be inferred as described in Section~\ref{sec:leakage-embedding} and are not yet included in knowledge $K$ are added.

\textbf{Results:}  
\begin{figure}
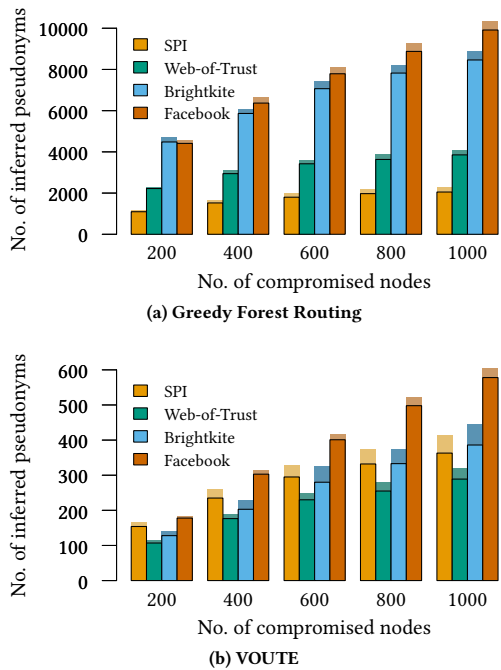

   \centering
   \subfloat[][Greedy Forest Routing]{
      \input{results/numPseud-gfr.tex}%
   } \\
   \subfloat[][VOUTE]{
      \input{results/numPseud-voute.tex}%
   }
   \caption{Mean number of inferred pseudonyms $N_p$ for different embedding algorithms, overlay graphs, numbers of compromised nodes and attacker behaviors. Saturated bars denote correct behavior and light bars on top denote increase due to deviating behavior. $99\%$ confidence intervals were omitted due to small size.}
   \label{fig:results-numPseud}
\end{figure}
Figure~\ref{fig:results-numPseud} shows the mean value for $N_p$ across the different graphs, attacker behaviors and number of compromised nodes.
Each point in Figure~\ref{fig:results-numPseud} shows to the mean value of $N_p$ over all 20 sets of malicious nodes, with 50 runs done per set.

By comparing the results from GFR with the results from VOUTE, it becomes apparent that the usage of enumeration indexes as coordinate allows malicious participants to infer roughly one order of magnitude more participating nodes and their coordinates than if random numbers are used.
While an adversary able to compromise 200 participants discovered on average around $152.8$ pseudonymous nodes on the SPI graph if VOUTE is used, they discovered around $1,090$ nodes if GFR is used.
On the Facebook graph, the number of inferred pseudonymous nodes increased from $174.2$ if VOUTE is used to $4,420$ if GFR is used.

The more elements a coordinate announced by a compromised node message has, the more likely it is that new pseudonymous nodes can be inferred.
As the average length of node coordinates decreases as the average hop distance to the root node decreases, we expected the number of inferred pseudonymous nodes $N_p$ to drop as the average shortest path length shrinks.
Contrary to our expectation, the mean value for $N_p$ on the Facebook graph was always the highest across all graphs, even though it has the lowest shortest path length on average among all graphs used for our study.
At the same time, the mean value for $N_p$ on the Web-of-Trust graph was always the lowest across all graphs for those runs where VOUTE is used as embedding algorithm, despite its high average shortest path length.
These results indicate that $N_p$ is more strongly affected by other properties, such as the graph's number of nodes, degree sequence as well as clustering.

By letting malicious nodes actively deviate from the correct behavior, the adversary is indeed able to infer more pseudonymous nodes than if malicious nodes operate correctly, although at a very limited scale for both embedding algorithms.
For example, on the Brightkite graph, the mean value for $N_p$ given 1000 compromised nodes increased by $4.6\%$ from $8,456$ to $8846.4$ inferred pseudonyms if $GFR$ is used when malicious nodes actively misbehaved.
In the runs where VOUTE was used, $N_p$ increased by $14.7\%$ from $387.1$ to $444.1$.
Similarly, on the Web-of-Trust graph, the mean value for $N_p$ increased by $5.8\%$ from $3,882.3$ to $4107.9$ for the runs with GFR and increased by $9.7\%$ from $289.4$ to $317.7$ for the runs with VOUTE.

\textbf{Summary of results:} Our study indicates that in overlay networks resembling social graphs, the usage of randomized coordinate elements reduces the number of participants that an attacker can infer from observed coordinates by at least one order of magnitude.
Contrary to our intuition, our results show that the average shortest path length is not the most decisive factor for the number of pseudonymous nodes the attacker is able to infer.
Furthermore, by letting malicious nodes only become leaf nodes, an attacker can increase the number of inferred pseudonyms by up to roughly $15\%$.

\section{Conclusion}
\label{sec:conclusion}
In this work, we analyzed the vulnerability of routing based on rooted spanning tree embeddings to inference attacks, in which adversaries aim to detect or even identify participants in an overlay network.
We showed that malicious participants can partially infer the structure of the encoded spanning tree from observed coordinates.
Furthermore, as most currently proposed algorithms use enumeration indexes as coordinate elements, malicious participants can additionally infer the coordinates of child nodes from each element.
To evaluate the feasibility of link inferences from observed data messages, we introduced the concept of a hypothetical overlay to represent the topological knowledge of an attacker, which takes potentially unknown links and participants into account.
Based on this concept, we showed that inference of links beyond the direct neighborhood of malicious nodes is not possible if the attacker cannot determine the originator of a message. 

Our simulation study indicates that in social graph-like networks, such as F2F overlays, the usage of random numbers as coordinate elements instead of enumeration indexes reduces the number of inferred tree nodes by more than one order of magnitude.
Furthermore, by letting malicious nodes keep their non-malicious neighbors from choosing them as parent in the spanning tree, an attacker can increase the number of inferred tree nodes by up to $15\%$.

From the proof regarding the inference of links from data messages, we identified the introduction of fake children nodes as a protection measure against link inferences in settings where the attacker may be aware of the coordinates of nearly all nodes.
Further research is needed to design such a countermeasure in way that an attacker cannot easily detect if a particular coordinate belongs to a fake or an actual child node.
Furthermore, further work is needed to investigate inferences that can be made if multiple such embeddings are formed in parallel and in the presence of network dynamics.

\bibliographystyle{amsplain}
\bibliography{articles}

\end{document}